\documentclass[aps,prl,preprint,groupedaddress]{revtex4-1}

\usepackage{graphicx}
\usepackage{array}

\usepackage{amssymb,amsmath}
\usepackage{subfig}
\usepackage{marvosym}
\usepackage{wasysym}
\usepackage{indentfirst}
\usepackage{tabularx}
\usepackage{indentfirst}
\usepackage{amsmath}
\usepackage{textcomp}
\usepackage{amssymb}
\usepackage{nicefrac}
\usepackage{marvosym}
\usepackage{yfonts}
\usepackage{notes2bib}
\usepackage{yfonts}
\usepackage{color}

\usepackage{svrsymbols}

\makeatletter
\newcommand\HUGE{\@setfontsize\normalsize{21}{21}}
\makeatother

\makeatletter
\newcommand\HUGGE{\@setfontsize\normalsize{18}{18}}
\makeatother

\makeatletter
\newcommand\LARGGE{\@setfontsize\normalsize{12}{12}}
\makeatother

\makeatletter
\newcommand\LARGGGE{\@setfontsize\normalsize{17}{17}}
\makeatother

\begin{document}


\title{Ideograms for Physics and Chemistry}




\author{Pablo Garc\'ia Risue\~no}
\email[For comments on conceptual issues and suggestions of new ideograms: ]{ pablo.risueno@chemie.uni-hamburg.de}
\affiliation{
Institut f\"ur Physikalische Chemie / Center for Ultrafast Imaging
Universit\"at Hamburg,
Grindelallee 117,
20146 Hamburg,
Germany 
}

\author{Apostolos Syropoulos}
\email[For technical respects: ]{ asyropoulos@yahoo.com}
\affiliation{Greek TeX Friends, 28 Oktovriou 366A, GR-67133 Xanthi, Greece }

\author{Nat\`alia Verg\'es}
\affiliation{Ars Didactica, C/ Major 2, 17850 Besal\'u, Girona, Spain}


\date{\today}

\indent{ }

\begin{abstract}

Ideograms (symbols that represent a word or idea) have great communicative value. 
They refer to concepts in a simple manner, easing the understanding of related ideas. 
Moreover, ideograms can simplify the often cumbersome notation used in the fields of Physics and physical Chemistry.
Nonetheless only a few specific ideograms for these fields -- like $\hbar$ and \AA$\ $-- have been defined to date.
In this work we propose that the scientific community follows the example of Mathematics 
--as well as that of oriental languages-- and bestows a more important role upon ideograms. 
To support this thesis we propose ideograms for essential concepts in Physics and Chemistry.
They are designed to be intuitive, and their goal is to make equations easier to read and understand.
Our symbols are included in a publicly available \LaTeX package ({\tt svrsymbols}). 

\indent{ }


\indent{ }


\indent{ }

\emph{The final publication is available at} {\tt link.springer.com}, \emph{DOI 10.1007/s10701-016-0044-5.}

\end{abstract}

\indent{ }

%






\maketitle

\section*{Introduction}

Ideograms are present in the communication codes that we use daily. 
All around the world, different kinds of ideograms are ubiquitous in daily life. 
Well known examples are traffic signs, musical notes, and commercial logos.
In written text, remarkable examples of ideograms are the integer signs
from 0 to 9 and other mathematical symbols, like $+,-,\%,\cdot,\times,\sqrt,\in $.
Symbols that represent a word, like @, \$ or \&, are common in texts written in English.
Ideograms are still more important in oriental languages: Japanese and Korean use mixed writing systems, with both phonetic (syllabic) and ideographic signs; in Japanese, almost all written sentences contain characters of both types, while Chinese, the language of more than one billion people, is mainly ideographic.

\indent{ }

For people whose mother tongue uses alphabetical and ideographic writing, ideograms convey a better and faster comprehension than the corresponding words written in phonetic characters \cite{shinamura}.
Moreover, concepts represented with ideograms are better identified and recalled \cite{arbuckle}. 
When reading a word written in alphabetical (or syllabic) characters, it is first necessary to identify every individual character (letter), then give them a phonetical value, and finally to associate the sound with the concept; in contrast, ideographic writing is simpler. It avoids the intermediate step, and so the meaning of the character (ideogram) can be \emph{directly} understood.
One can intuitively gauge the advantage of ideograms by merely comparing the ease of reading numbers
(1, 2, 3 ...) and their written names (one, two, three ...); this phenomenon was thoroughly studied in \cite{besner}.
The expressive power of ideograms is also illustrated by the dramatic fact that there exist people with neurological 
injuries who cannot read texts written with phonetic signs, but they can read and understand well if the text is written with ideograms \cite{sasanuma}. 

\indent{ }

Signs are often classified into \emph{symbols}, \emph{indices} or \emph{icons} depending on its similarity
with the object they refer to. \emph{Symbols} are purely abstract, while \emph{indices} are somewhat connected 
to the object, and \emph{icons} resemble the object. 
The proximity of signs and their signified objects helps to recall the latter in a more vivid way. 
Umberto Eco, a celebrated expert in semiotics, stated that an icon can acquire primacy over its signified due to familiarity:
''At a certain point the iconic representation, however stylized it may be, appears to be more true than the real experience, and people begin to look at things through the glasses of iconic convention'' \cite{ecooo}. 
 This statement is exemplified e.g. by experiments that prove that human mind identifies the concept 'hand' on a cartoon of it more quickly than on a photography of a hand \cite{ryans}.
 Albert Einstein, frequently considered the best scientist ever, stated that he did not think in terms of words, but of symbols \cite{gruberbook}.

\indent{ }

We propose that the power of ideograms should be more often utilized by the scientific community. 
The fact that this community has been mostly composed by westerners has probably precluded this possibility so far.
However, the current complexity of equations suggests a universal notation which is as intuitive and univocal as possible. 
Despite the close relationship between Mathematics and Physics, ideograms are ubiquitous in Mathematics, but
shyly used in Physics. Examples of used ideograms are $\hbar$ (reduced Planck constant), \AA$\ $ (angstrom)  or symbols for particles used
in particle Physics (see appendix).
We wish to advocate for the definition of new ideograms for the most important concepts of Physics and Chemistry.
The form and the meaning of such ideograms should be agreed by the community, e.g.
in specialized conferences as those dedicated to the definition of units
\cite{note1}.
These ideograms should be 
simple and intuitive, and they should
clearly point to specific concepts to avoid any confusion to their meaning.

\indent{ }

In order to boost the trend of using ideograms in Science, we present below examples for the representation of
basic concepts of Physics and Chemistry.
All symbols presented out of the appendix are original, except those of atom, spin, hole and the first symbols for particles
on table \ref{tab:2} (the appendix also contains symbols traditionally used in particle Physics).  
We welcome suggestions for additional new symbols in any field of Physics and Chemistry.

\indent{ }

We have tried to design ideograms which are intuitive and easy to remember. To this end 
we based many of our designs on letters of the Latin alphabet and we tried to give them an iconic or index-character.
For example, we added wiggly lines at the ideograms for 'photon' or 'phonon', and the symbol for 'graphene' 
is nearly an hexagon. To ease their identification, some of our designs contain the initial letter of the 
word that they represent. The chosen font in most cases is {\tt cmmi12} or {\tt cmr10}, which are
widely used for writing equations.

\indent{ }

\indent{ }

\section*{Ideograms for general concepts}

The six signs on table \ref{tab:1} are generic, they can be used in any discipline.
\begin{center}
\begin{table}[h!]
\renewcommand{\arraystretch}{1.7}
\bgroup
\def\arraystretch{1.5}
\begin{tabular}{rcrcrc}
{\LARGGE Experimental } & { \HUGE $  \experimentalsym \qquad \quad $ } 
&
  {\LARGGE Error } & {\HUGE $  \errorsym \qquad \quad$} 
&
 {\LARGGE Method } & {\HUGE $  \method$} 
\\
 {\LARGGE Assumption }  & {\HUGE $\assumption \qquad \quad$} 
&
 {\LARGGE Example }  & {\HUGE $  \svrexample  \qquad \quad$} 
&
  {\LARGGE Reference } &  {\HUGE $ \reference$} 
\end{tabular}
\egroup
\caption{\emph{\small Ideograms for generic concepts.}}
\label{tab:1}
\end{table}
\end{center}
The use of $\errorsym$ makes it clear that the represented quantity means 'error'; the use of $E$ or $e$ 
instead would conflict with 'energy', '2.71828..', 'charge of the electron', etc. '$Err$' has three 
letters which makes it slower to read and could be misinterpreted as the product of three quantities.
The use of $\experimentalsym$ instead of 'exp' can avoid the confusion with 'exponential'.
Naturally every scientist does understand the concepts behind traditional notation (like '$Err$' or 'exp'); 
but the need of making a short-time reflection on the meaning of these characters
distracts from the main goal of understanding the physical meaning of the equation in question.
Note that $\experimentalsym$ is different to the Cyrillic character 'Zhe'.

\indent{ }

The sign of assumption ($\assumption$) was chosen to make it stand out when browsing a page. 
Frequently, is difficult to locate the underlying assumptions in many scientific papers and a 
scientist may spend an inordinate amount of time before discovering that the paper is not 
relevant for his own line of investigation. If a big $\assumption$ sign appears whenever an 
assumption is made, the reader will immediately be able to determine if the paper is suitable for his research.
We recommend the use of the notation $\assumption$1, $\assumption$2, $\ldots$, $\assumption$$n$ for the 1st, 2nd, $\ldots$, $n$-th assumptions, for the sake of future references.

\indent{ }

\indent{ }

\section*{Ideograms for particles and quasiparticles}

The signs in the first two rows of table \ref{tab:2} (among others) are used in particle 
Physics, but to the best of our knowledge they have never included into a \LaTeX package.
We designed them to be more elegant than their {\tt math mode} counterparts ($e^-$, $p^+$, $\ldots$).
The existence of ideograms for particles like 'electron' makes it advisable to have at our disposal specific signs for other widely used concepts in Physics, such as
'photon' or 'phonon'.

\begin{center}
\begin{table}[h!]
\renewcommand{\arraystretch}{1.7}
\bgroup
\def\arraystretch{1.5}
\begin{tabular}{rcrcrc}
 {\LARGGE Electron } & {\HUGE $\electron \qquad \quad$} 
&
 {\LARGGE Proton }  & {\HUGE $\proton \qquad \quad$} 
&
 {\LARGGE Neutron }  & {\HUGE $\neutron \  $} 
\\
 {\LARGGE Positron }  & {\HUGE $\positron \qquad \quad$} 
&
 {\LARGGE Antiproton } &  {\HUGE $\antiproton \qquad \quad$} 
&
 {\LARGGE Antineutron } &  {\HUGE $\antineutron \ $} 
\\
 {\LARGGE Fermion }  & {\HUGE $\fermion  \qquad \quad$} 
&
 {\LARGGE Boson }  & {\HUGE $ \boson  \qquad \quad$} 
&
 {\LARGGE Anyon } &  {\HUGE $\anyon  \ $} 
\\
 {\LARGGE Photon }  & {\HUGE $ \svrphoton, \ \varphoton  \qquad \quad$  } 
&
 {\LARGGE Phonon }  & {\HUGE $\phonon \qquad \quad$} 
&
 {\LARGGE Polaron }  & {\HUGE $\polaron \ $} 
\\
 {\LARGGE Polariton  } &  {\HUGE $ \polariton \qquad \quad$} 
&
 {\LARGGE Hole }  & {\HUGE $\hole \qquad \quad$} 
&
 {\LARGGE Magnon }  & {\HUGE $ \magnon  \ $} 
\\
 {\LARGGE Exciton }  & {\HUGE $\exciton \qquad \quad$} 
&
 {\LARGGE Plasmon }  & {\HUGE $\plasmon \qquad \quad$} 
&
  {\LARGGE Tachyon}  & {\HUGE $ \tachyon  \ $}
\\
 {\LARGGE J/$\Psi$ meson }  & {\HUGE $\Jpsimeson \qquad \quad$} 
&
 {\LARGGE Higgs boson }  & {\HUGE $\Higgsboson \qquad \quad$} 
&
  {\LARGGE Graviton}  & {\HUGE $ \graviton  \ $}
\end{tabular}
\egroup
\caption{\emph{\small Ideograms for particles and quasiparticles.}}
\label{tab:2}
\end{table}
\end{center}
The wiggly lines in the ideograms of table \ref{tab:2} emphasize on the wave nature of (quasi)particles. 
The signs of 'photon' and 'phonon' are based on the '$f$' and '$F$' characters due to the equivalence between the 'ph' and 'f' sounds.
The symbol of 'magnon' is similar to that of 'spin' (see table \ref{tab:4}), but suggesting a wave propagation. 
The ideogram for 'graviton' resembles a spiral galaxy.
The sign of '$\times$' within the 'polariton' represents an excitation in the material.
The symbols of \emph{boson} and \emph{fermion} require some further explanation. 
During the middle ages, a typical topic of philosophical discussion was the question: 
''How many angels can stand on the point of a pin?''. The accepted answer was ''An infinity'' due to the 
immaterial character of angels. The symbol of 'boson' resembles a $B$ formed with the wings of an angel
(because bosons, like angels, do not obstruct each other). In contrast, the symbol of 'fermion' resembles an $F$
formed with the trident of a demon (designing atomic weapons as F. did is considered by many to be somewhat devilish).
%
%
\indent{ }

We also include many symbols that are used at the field of particle Physics into the {\tt svrsymbols} package. They can be viewed in the appendix.

\indent{ }

\indent{ }

\section*{Ideograms for concepts of chemistry} 

The symbols presented on table \ref{tab:3} are a suggestion of some ideograms for basic concepts of Chemistry. 
We suggest the design of many others, pointed by the Chemistry community.


\begin{center}
\begin{table}[ht!]
\renewcommand{\arraystretch}{1.7}
\bgroup
\def\arraystretch{1.5}
\begin{tabular}{rcrcrc}
{\LARGGE Chemical bond }  & {\HUGE $ \bond \quad$} 
&
{\LARGGE Ionic bond }  & {\HUGE $ \ionicbond  \quad$} 
&
{\LARGGE Metallic bond } &  {\HUGE $ \metalbond \ $} 
\\
{\LARGGE Single covalent bond } &  {\HUGE $\covbond \quad$} 
&
{\LARGGE Double cov. bond }  & {\HUGE $ \doublecovbond  \quad$} 
&
{\LARGGE Triple cov. bond }  & {\HUGE $ \triplecovbond \   $} 
\\
{\LARGGE Hydrogen bond }  & {\HUGE $ \hbond \quad$} 
&
{\LARGGE Water } &  {\HUGE $\water \quad$} 
&
{\LARGGE Protein }  & {\HUGE $\protein \ $} 
\\
{\LARGGE Adsorbent }  & {\HUGE $\adsorbent \quad$} 
&
{\LARGGE Adsorbate } &  {\HUGE $ \adsorbate \quad$} 
& 
\end{tabular}
\egroup
\caption{\emph{\small Ideograms for basic concepts of Chemistry.}}
\label{tab:3}
\end{table}
\end{center}

The signs for bonds on table \ref{tab:3} have the shape of a double harpoon to stress the nature of bonds as joining
objects. The different kinds of bonds are specified by their initials to ease their identification. 
The ideogram for 'protein' is a P-shaped alpha-helix, while the 'adsorbate' is a modified A and
the sign of 'adsorbent' emphasizes on its capability to join to the adsorbate.

\indent{ }

\indent{ }

\section*{Ideograms for other systems and phenomena}

The symbols presented on table
\ref{tab:4} correspond to a variety of concepts which are commonly used in Physics. 
\begin{center}
\begin{table}[h!]
\renewcommand{\arraystretch}{1.7}
\bgroup
\def\arraystretch{1.5}
\begin{tabular}{rlrcrc}
{\LARGGE Atom }  & {\HUGE $\ \atom $} 
&
{\LARGGE Nucleus }  & {\HUGE $\nucleus \quad $} 
&
{\LARGGE Ion } &  {\HUGE $\ion $} 
\\
{\LARGGE Spin (up, down) } &  {\HUGE $ \spin \ ( \spinup ,  \spindown ) $} 
&
{\LARGGE Orbit  }  & {\HUGE $ \orbit \quad$} 
&
{\LARGGE Graphene }  & {\HUGE $\graphene  $} 
\\
{\LARGGE Dipole   }  & {\HUGE  $ \ \  \dipole  $} 
&
{\LARGGE Quadrupole   }  & {\HUGE $\quadrupole \quad $} 
&
{\LARGGE Surface } &  {\HUGE $\surface $} 
\\
{\LARGGE Maxwell distribution} &  {\HUGE $\ \maxwellDistrib   $} 
&
{\LARGGE Fermi distribution}  & {\HUGE $ \fermiDistrib \quad  $} 
&
{\LARGGE Bose distribution} &  {\HUGE $ \boseDistrib  $} 
\\
{\LARGGE Solid }  & {\HUGE $\  \solid  $ } 
&
{\LARGGE Conductivity } &  {\HUGE $ \conductivity \quad  $} 
&
{\LARGGE Resistivity } &  {\HUGE $ \resistivity   $} 
\\
{\LARGGE Internal } &  {\HUGE $ \internalsym    $} 
&
{\LARGGE External } &  {\HUGE $\ \externalsym  \quad  $} 
&
{\LARGGE Interaction } &  {\HUGE $\ \interaction    $} 
\end{tabular}
\egroup
\caption{\emph{\small  Ideograms for several concepts of Physics.}}
\label{tab:4}
\end{table}
\end{center}
The ideogram of 'dipole' on table
\ref{tab:4} is made of a $d$ and a $p$ resembling two charged spheres at the terminals of a segment;
the ideogram of 'quadrupole' is made of two symbols of 'dipole'. The symbol of 'surface' is based on an S, 
while that of 'graphene' is based on a G (with a hexagonal shape, like the primitive cell of graphene). The symbol of
conductivity is wheel-shaped to stress on the mobility of charges, and the ideogram of 'resistivity' tries to 
hint the obstruction to the motion.

%
%



\indent{  }

\indent{ }

Let us see some examples on how the proposed ideograms simplify the notation. A typical expression for a 
Hamiltonian depending on electronic and phononic parts is
$H \  = \  H_{el} + H_{ph}  +   H_{el-ph}$.
With our simplified notation, its expression would be
$H \  = \  H_{\electron}  +  H_{\phonon} +  H_{\electron\phonon}$, or
$H \  = \  H_{\electron}  +  H_{\phonon} +  H_{\interaction}$.
The example of a table where the results and errors of different calculation methods are 
compared with experimental results looks as follows:
\begin{center}
\begin{table}[h!]
\begin{tabular}{ccc}
\begin{tabular}{|c|c|c|c|c|}
\hline
$Method$ 1 & $Err. meth. \ 1$ & $Method$ 2 & $Err. meth. \ 2$ & $Exp.$ \\
\hline
\hline
1.01 & 0.12 & 1.12 & 0.03 & 1.15 \\
2.56 & 0.14 & 2.28 & 0.02 & 2.24 \\
\hline
\end{tabular}
\begin{tabular}{ccc}
& & \\
& & \\
\end{tabular}
\begin{tabular}{|c|c|c|c|c|}
\hline
$\method 1$  & $\errorsym_{\method  1}$ & $\method 2$  & $\errorsym_{\method 2}$ & $\ \experimentalsym \ $ \\
\hline
\hline
1.01 & 0.12 & 1.12 & 0.03 & 1.15 \\
2.56 & 0.14 & 2.28 & 0.02 & 2.24 \\
\hline
\end{tabular}
\end{tabular}
\caption{\emph{\small  Comparison of data presentation without and with the use of ideograms.}}
\end{table}
\end{center}
In these examples it is clear that the ideographic notation is simpler, and it does not require 
the reader to decode the phonetical value and the meaning of a set of letters because the meaning of an ideogram
is univocal. Such examples illustrate the usefulness and the communication capabilities of ideograms in Physics and Chemistry. 

\indent{ }

We expect that the presented ideograms are themselves useful for the scientific community,
and that they promote the use of specifically-devised ideograms for basic concepts of Physics and Chemistry, as it is customary and useful in both oriental languages and Mathematics.

\indent{  }

\section{Acknowledgments}
\begin{acknowledgments}
We are very grateful to Carl Christian K. Mikkelsen (University of Ume\aa, Sweden) for his help and advice.
\end{acknowledgments}

\section{Author contributions}
P.G.R. proposed the idea of developing and using further ideograms and made by hand the corresponding designs; both N.V. and A.S. worked as graphical designers; A.S. wrote the \LaTeX package that includes the ghlyphs.

\indent{  }

\indent{  }

\indent{  }

\clearpage

\appendix\label{appendix}

\section{Appendix: Table of glyphs of the {\tt svrsymbols} package}

The {\tt SVRsymbols} font contains new ideograms for use in Science.
It also contains a standard presentation for ideograms which are traditionally used in particle Physics
(e.g. $\electron$, $\proton$) but which have not previously been included in \LaTeX. 
We included them into the {\tt svrsymbols} package
so that there is a standard presentation that boosts the use of ideograms in Science.
For some concepts there existed a sign but we propose an alternative ideogram
to simplify the notation ($J/\Psi \rightarrow \Jpsimeson$) or
to avoid confusion 
(e.g. $\gamma \rightarrow \svrphoton $; $H^0 \rightarrow \Higgsboson$; $G \rightarrow \graviton$)\cite{note2}.
We suggest to gradually change the rest of 
characters used for particles (e.g. $c$, $d$, $g$) by specifically-devised ideograms.
The novel symbols have been designed to be intuitive and easy to identify and to remember. 
The {\tt SVRsymbols} package has no options and it provides an interface to the font.
It defines commands for use in math mode. The commands and the resulting symbols are shown in the table below.
The package can be downloaded from {\tt http://ctan.org/pkg/svrsymbols} and freely used
if the present letter is cited.

\begin{center}
\renewcommand{\arraystretch}{2}
\bgroup
\def\arraystretch{1.5}
\begin{tabular}{rlrl}
 {\LARGGE Adsorbent ({\tt \textbackslash adsorbent})} &  {\HUGE $\ \ \  \adsorbent$} 
&
 {\LARGGE Adsorbate ({\tt \textbackslash  adsorbate})}  & {\HUGE $ \ \ \   \adsorbate$} 
\\
 {\LARGGE Antimuon ({\tt \textbackslash  antimuon})}  & {\HUGE $\ \ \   \antimuon$}
&
 {\LARGGE Antineutrino ({\tt \textbackslash antineutrino})}  & {\HUGE $\ \ \  \antineutrino$}
\\
 {\LARGGE Antineutron ({\tt \textbackslash antineutron})}  & {\HUGE $\ \ \  \antineutron$} 
&
 {\LARGGE Antiproton ({\tt \textbackslash antiproton})} &  {\HUGE $\ \ \  \antiproton$} 
\\
 {\LARGGE Antiquark ({\tt \textbackslash antiquark})}  & {\HUGE $\ \ \  \antiquark$} 
&
 {\LARGGE Anyon ({\tt \textbackslash  anyon})}  & {\HUGE $\ \ \  \anyon $} 
\\
 {\LARGGE Assumption ({\tt \textbackslash assumption})} & {\HUGE $\ \  \assumption$} 
&
 {\LARGGE Assumption ({\tt \textbackslash bigassumption})} & {\HUGE $\ \  \bigassumption$} 
\\
 {\LARGGE Assumption ({\tt \textbackslash biggassumption})} & {\HUGE $\ \  \biggassumption$} 
&
 {\LARGGE Assumption ({\tt \textbackslash Bigassumption})} & {\HUGE $\ \  \Bigassumption$} 
\\
 {\LARGGE Atom ({\tt \textbackslash  atom})}  & {\HUGE $\ \ \  \atom$} 
& 
 {\LARGGE $B^+$ meson ({\tt \textbackslash Bmesonplus})}  & {\HUGE $\  \ \Bmesonplus \  $}
\\
 {\LARGGE $B^0$ meson ({\tt \textbackslash Bmesonnull})}  & {\HUGE $ \  \  \Bmesonnull \  $}
&
 {\LARGGE $B^-$ meson ({\tt \textbackslash Bmesonminus})}  & {\HUGE $\ \  \Bmesonminus  \  $}
\\
 {\LARGGE Bose distribution ({\tt \textbackslash  boseDistrib})}  & {\HUGE $\ \ \  \boseDistrib $} 
&
 {\LARGGE Boson ({\tt \textbackslash  boson})} &  {\HUGE $\ \ \  \boson $} 
\\
 {\LARGGE Bottom antiquark ({\tt \textbackslash antiquarkb})}  & {\HUGE $\ \ \  \antiquarkb$}
&
 {\LARGGE Bottom quark ({\tt \textbackslash  quarkb})}  & {\HUGE $\ \ \  \quarkb$ }
 \\
 \\
 \end{tabular}

\begin{tabular}{rlrl}
 {\LARGGE Charm antiquark ({\tt \textbackslash antiquarkc})}  & {\HUGE $\ \ \  \antiquarkc$} 
&
 {\LARGGE Charm quark ({\tt \textbackslash  quarkc})}  & {\HUGE $\ \ \  \quarkc$ } 
\\
 {\LARGGE Chemical bond ({\tt \textbackslash  bond})} &  {\HUGE $\ \  \bond$} 
&
 {\LARGGE Conductivity ({\tt \textbackslash  conductivity}) }  & {\HUGE $\ \    \conductivity $} 
\\
 {\LARGGE $D^+$ meson ({\tt \textbackslash Dmesonplus})}  & {\HUGE $\ \  \Dmesonplus \  $}
&
 {\LARGGE $D^0$ meson ({\tt \textbackslash Dmesonnull})}  & {\HUGE $  \  \  \Dmesonnull \  $}
\\
 {\LARGGE $D^-$ meson ({\tt \textbackslash Dmesonminus})}  & {\HUGE $\  \ \Dmesonminus  \  $}
&
{\LARGGE Dipole  ({\tt \textbackslash  dipole}) } &  {\HUGE  $\ \ \  \dipole$} 
\\
 {\LARGGE Dbl. cov. bond ({\tt \textbackslash doublecovbond})}  & {\HUGE $ \  \doublecovbond$} 
&
 {\LARGGE Down antiquark ({\tt \textbackslash antiquarkd})}  & {\HUGE $\ \ \  \antiquarkd$} 
\\
{\LARGGE Down quark ({\tt \textbackslash  quarkd})}  & {\HUGE $\ \ \  \quarkd $} 
&
 {\LARGGE Electron ({\tt \textbackslash electron})}  & {\HUGE $\ \ \  \electron$} 
\\
 {\LARGGE Error ({\tt \textbackslash errorsym})} & {\HUGE $\  \  \errorsym$} 
&
 {\LARGGE Example ({\tt \textbackslash svrexample})}  & {\HUGE $\quad \svrexample  $} 
\\
 {\LARGGE Exciton ({\tt \textbackslash  exciton})} &  {\HUGE $ \  \exciton$} 
&
{\LARGGE Experimental ({\tt \textbackslash experimentalsym})} & { \HUGE $\  \   \experimentalsym \quad \ \ $ } 
\\
 {\LARGGE External ({\tt \textbackslash externalsym})}  & {\HUGE $\ \  \externalsym  $} 
&
{\LARGGE Fermi distribution ({\tt \textbackslash  fermiDistrib})}  & {\HUGE $\ \ \  \fermiDistrib$} 
\\
 {\LARGGE Fermion ({\tt \textbackslash  fermion})}  & {\HUGE $\  \  \fermion $} 
&
 {\LARGGE Gluon ({\tt \textbackslash Gluon})}  & {\HUGE $\ \ \ \Gluon  \  $}
\\
{\LARGGE Graphene ({\tt \textbackslash  graphene})} &  {\HUGE $\    \graphene$} 
&
 {\LARGGE Graviton ({\tt \textbackslash Graviton})}  & {\HUGE $\ \ \  \graviton  $}
\\
 {\LARGGE Higgs boson ({\tt \textbackslash Higgsboson})}  & {\HUGE $\  \Higgsboson  $}
&
 {\LARGGE Hole ({\tt \textbackslash  hole})} &  {\HUGE $\ \ \  \hole$} 
\\
{\LARGGE Hydrogen bond ({\tt \textbackslash  hbond})}  & {\HUGE $\   \hbond$} 
&
 {\LARGGE Interaction ({\tt \textbackslash interaction})} &  {\HUGE $\ \ \  \interaction  $} 
\\
 {\LARGGE Internal ({\tt \textbackslash internalsym})}  & {\HUGE $\   \internalsym  $} 
&
{\LARGGE Ion ({\tt \textbackslash  ion})}  & {\HUGE $\ \ \  \ion$} 
\\
{\LARGGE Ionic bond ({\tt \textbackslash  ionicbond})} &  {\HUGE $\   \ionicbond $} 
&
 {\LARGGE J/$\Psi$ meson ({\tt \textbackslash Jpsimeson})}  & {\HUGE $\ \ \  \Jpsimeson  $}
\\
 {\LARGGE Kaons ({\tt \textbackslash Kaonplus})}  & {\HUGE $\ \Dmesonplus  \  $}
&
 {\LARGGE Kaons ({\tt \textbackslash Kaonnull})}  & {\HUGE $\ \ \  \Dmesonnull  \  $}
 \\
  {\LARGGE Kaons ({\tt  \textbackslash Kaonminus})}  & {\HUGE $\  \Dmesonminus  \  $}
&
 {\LARGGE Magnon ({\tt \textbackslash  magnon})}  & {\HUGE $\ \ \  \magnon $} 
 \\
  {\LARGGE Maxwell distrib. ({\tt \textbackslash  maxwellDistrib})}  & {\HUGE $\   \maxwellDistrib  $} 
&
{\LARGGE Metallic bond ({\tt \textbackslash  metalbond})}  & {\HUGE $\ \ \  \metalbond $} 
\\
 {\LARGGE Method ({\tt \textbackslash method})} & {\HUGE $\   \method$} 
&
 {\LARGGE Muon ({\tt \textbackslash  muon})}  & {\HUGE $\ \ \  \muon$ }
\\
\\
\end{tabular}

\begin{table}[h!]
\begin{tabular}{rlrl}
 {\LARGGE Neutrino ({\tt \textbackslash neutrino})}  & {\HUGE $\ \ \   \neutrino$ }
&
 {\LARGGE Neutron ({\tt \textbackslash neutron})}  & {\HUGE $\ \ \  \neutron$} 
\\
{\LARGGE Nucleus ({\tt \textbackslash  nucleus})}  & {\HUGE $\  \  \nucleus$} 
&
{\LARGGE Orbit ({\tt \textbackslash  orbit}) }  & {\HUGE $\ \  \orbit$} 
\\
{\LARGGE Phi mesons ({\tt \textbackslash phimeson}) }  & {\HUGE $\ \  \  \phimeson $} 
&
{\LARGGE Phi mesons ({\tt  \textbackslash phimesonnull}) }  & {\HUGE $\  \ \ \phimesonnull $} 
\\
 {\LARGGE Phonon ({\tt \textbackslash  phonon})} &  {\HUGE $\  \  \phonon$} 
&
 {\LARGGE Photon ({\tt \textbackslash svrphoton,\textbackslash varphoton})}  & {\HUGE $\   \svrphoton ,  \varphoton$  } 
\\
 {\LARGGE Pion plus ({\tt \textbackslash pionplus})}  & {\HUGE $\ \ \pionplus  \  $}
&
 {\LARGGE Pion 0 ({\tt \textbackslash pionnnull})}  & {\HUGE $\ \ \pionnull  \  $}
\\
 {\LARGGE Pion minus ({\tt \textbackslash pionminus})}  & {\HUGE $\  \ \pionminus  \  $}
&
{\LARGGE Plasmon ({\tt \textbackslash  plasmon})} &  {\HUGE $\    \plasmon$} 
\\
{\LARGGE Polariton  ({\tt \textbackslash  polariton})}  & {\HUGE $\ \ \polariton$} 
&
{\LARGGE Polaron ({\tt \textbackslash  polaron})}  & {\HUGE $ \  \polaron$} 
\\
 {\LARGGE Positron ({\tt \textbackslash positron})} &  {\HUGE $\ \ \positron$} 
&
{\LARGGE Protein ({\tt \textbackslash  protein})}  & {\HUGE $\ \ \  \protein$} 
\\
 {\LARGGE Proton ({\tt \textbackslash proton})} &  {\HUGE $\   \ \proton$} 
&
 {\LARGGE Quadrupole  ({\tt \textbackslash  quadrupole}) } &  {\HUGE $\ \   \quadrupole$} 
 \\
  {\LARGGE Quark ({\tt \textbackslash quark})}  & {\HUGE $\ \   \quark$ } 
&
 {\LARGGE Reference ({\tt \textbackslash reference})}  &  {\HUGE $\ \ \  \reference$} 
\\
{\LARGGE Resistivity ({\tt \textbackslash  resistivity}) }  & {\HUGE $ \ \  \resistivity $} 
&
 {\LARGGE Rho meson plus ({\tt \textbackslash rhomesonplus})}  & {\HUGE $\ \ \  \rhomesonplus  \  $}
\\
 {\LARGGE Rho meson 0 ({\tt  \textbackslash rhomesonnull})}  & {\HUGE $\ \ \rhomesonnull \  $}
&
 {\LARGGE Rho meson minus ({\tt  \textbackslash rhomesonminus})}  & {\HUGE $\ \ \ \rhomesonminus  \  $}
\\
{\LARGGE Single covalent bond ({\tt \textbackslash  covbond})}  & {\HUGE $\    \covbond$} 
&
{\LARGGE Solid ({\tt \textbackslash  solid})}  & {\HUGE $\ \  \solid$ } 
\\
 {\LARGGE Spin  ({\tt \textbackslash  spin})} & {\HUGE $\ \   \spin $} 
&
 {\LARGGE Spin down ({\tt \textbackslash  spindown})} & {\HUGE $\ \  \ \spindown$} 
\\
 {\LARGGE Spin up ({\tt \textbackslash  spinup})} & {\HUGE $\ \   \spinup$} 
&
 {\LARGGE Strange antiquark ({\tt \textbackslash antiquarks})} &  {\HUGE $\  \ \  \antiquarks$} 
\\
 {\LARGGE Strange quark ({\tt \textbackslash  quarks})} &  {\HUGE $\ \   \quarks$ } 
&
 {\LARGGE Surface ({\tt \textbackslash surface})}  & {\HUGE $\  \  \surface $} 
 \\
 {\LARGGE $T^+$ meson ({\tt \textbackslash Tmesonplus})}  & {\HUGE $\ \Tmesonplus \  $}
&
 {\LARGGE $T^0$ meson ({\tt \textbackslash Tmesonnull})}  & {\HUGE $ \ \  \Tmesonnull \  $}
\\
 {\LARGGE $T^-$ meson ({\tt \textbackslash Tmesonminus})}  & {\HUGE $\  \Tmesonminus  \  $}
&
 {\LARGGE Tachyon ({\tt \textbackslash tachyon})}  &  {\HUGE $ \ \ \tachyon  \ $}
 \\
 {\LARGGE Tau lepton plus ({\tt \textbackslash  tauleptonplus})} &  {\HUGE $\ \ \tauleptonplus$}
 &
 {\LARGGE Tau lepton minus ({\tt  \textbackslash tauleptonminus})} &  {\HUGE $\ \ \tauleptonminus$}
\end{tabular}
\end{table}

\begin{table}[h!]
\begin{tabular}{rlrl}
  {\LARGGE Top antiquark ({\tt \textbackslash antiquarkt})} &  {\HUGE $\ \ \  \antiquarkt$} 
 & 
  {\LARGGE Top quark ({\tt \textbackslash  quarkt})} &  {\HUGE $\  \   \quarkt $} 
\\
{\LARGGE Triple cov. bond ({\tt \textbackslash  triplecovbond})}  & {\HUGE $ \triplecovbond$} 
&
 {\LARGGE Up antiquark ({\tt \textbackslash  antiquarku})}  & {\HUGE $\ \  \antiquarku$} 
\\
 {\LARGGE Up quark ({\tt \textbackslash  quarku})}  & {\HUGE $\    \ \quarku$ } 
&
 {\LARGGE Upsilon meson ({\tt \textbackslash Upsilonmeson })}  & {\HUGE $\ \   \Upsilonmeson  $}
\\
{\LARGGE Water ({\tt \textbackslash  water})}  & {\HUGE $\ \  \water$} 
&
 {\LARGGE W bosons ({\tt \textbackslash Wboson})}  & {\HUGE $\ \Wboson  \  $}
\\
 {\LARGGE W bosons ({\tt  \textbackslash Wbosonplus})}  & {\HUGE $\ \Wbosonplus \  $}
&
 {\LARGGE W bosons ({\tt \textbackslash Wbosonminus})}  & {\HUGE $\  \Wbosonminus  \  $}
\\
{\LARGGE Z boson ({\tt \textbackslash  Zboson})}  & {\HUGE $\  \  \Zboson$} 
\end{tabular}
\caption{\emph{\small Ideograms of the {\tt svrsymbols} \LaTeX package with their corresponding commands.}}
\end{table}
\egroup
\end{center}

\indent{  }

%
%
%

\indent{  }

\indent{  }


\bibliography{biblio}

\begin{thebibliography}{9}%
\makeatletter
\providecommand \@ifxundefined [1]{%
 \@ifx{#1\undefined}
}%
\providecommand \@ifnum [1]{%
 \ifnum #1\expandafter \@firstoftwo
 \else \expandafter \@secondoftwo
 \fi
}%
\providecommand \@ifx [1]{%
 \ifx #1\expandafter \@firstoftwo
 \else \expandafter \@secondoftwo
 \fi
}%
\providecommand \natexlab [1]{#1}%
\providecommand \enquote  [1]{``#1''}%
\providecommand \bibnamefont  [1]{#1}%
\providecommand \bibfnamefont [1]{#1}%
\providecommand \citenamefont [1]{#1}%
\providecommand \href@noop [0]{\@secondoftwo}%
\providecommand \href [0]{\begingroup \@sanitize@url \@href}%
\providecommand \@href[1]{\@@startlink{#1}\@@href}%
\providecommand \@@href[1]{\endgroup#1\@@endlink}%
\providecommand \@sanitize@url [0]{\catcode `\\12\catcode `\$12\catcode
  `\&12\catcode `\#12\catcode `\^12\catcode `\_12\catcode `\%12\relax}%
\providecommand \@@startlink[1]{}%
\providecommand \@@endlink[0]{}%
\providecommand \url  [0]{\begingroup\@sanitize@url \@url }%
\providecommand \@url [1]{\endgroup\@href {#1}{\urlprefix }}%
\providecommand \urlprefix  [0]{URL }%
\providecommand \Eprint [0]{\href }%
\providecommand \doibase [0]{http://dx.doi.org/}%
\providecommand \selectlanguage [0]{\@gobble}%
\providecommand \bibinfo  [0]{\@secondoftwo}%
\providecommand \bibfield  [0]{\@secondoftwo}%
\providecommand \translation [1]{[#1]}%
\providecommand \BibitemOpen [0]{}%
\providecommand \bibitemStop [0]{}%
\providecommand \bibitemNoStop [0]{.\EOS\space}%
\providecommand \EOS [0]{\spacefactor3000\relax}%
\providecommand \BibitemShut  [1]{\csname bibitem#1\endcsname}%
\let\auto@bib@innerbib\@empty
\bibitem [{\citenamefont {Shinamura}(1987)}]{shinamura}%
  \BibitemOpen
  \bibfield  {author} {\bibinfo {author} {\bibfnamefont {A.~P.}\ \bibnamefont
  {Shinamura}},\ }\href@noop {} {\bibfield  {journal} {\bibinfo  {journal} {The
  American Journal of Psychology}\ }\textbf {\bibinfo {volume} {100}},\
  \bibinfo {pages} {15} (\bibinfo {year} {1987})}\BibitemShut {NoStop}%
\bibitem [{\citenamefont {Park}\ and\ \citenamefont
  {Arbuckle}(1977)}]{arbuckle}%
  \BibitemOpen
  \bibfield  {author} {\bibinfo {author} {\bibfnamefont {S.}~\bibnamefont
  {Park}}\ and\ \bibinfo {author} {\bibfnamefont {T.~Y.}\ \bibnamefont
  {Arbuckle}},\ }\href@noop {} {\bibfield  {journal} {\bibinfo  {journal}
  {Journal of Experimental Psychology: Human Learning and Memory}\ }\textbf
  {\bibinfo {volume} {3(6)}},\ \bibinfo {pages} {631} (\bibinfo {year}
  {1977})}\BibitemShut {NoStop}%
\bibitem [{\citenamefont {Besner}\ and\ \citenamefont
  {Coltheart}(1979)}]{besner}%
  \BibitemOpen
  \bibfield  {author} {\bibinfo {author} {\bibfnamefont {D.}~\bibnamefont
  {Besner}}\ and\ \bibinfo {author} {\bibfnamefont {M.}~\bibnamefont
  {Coltheart}},\ }\href@noop {} {\bibfield  {journal} {\bibinfo  {journal}
  {Neuropsychologia}\ }\textbf {\bibinfo {volume} {17}},\ \bibinfo {pages}
  {467} (\bibinfo {year} {1979})}\BibitemShut {NoStop}%
\bibitem [{\citenamefont {Sasanuma}(1975)}]{sasanuma}%
  \BibitemOpen
  \bibfield  {author} {\bibinfo {author} {\bibfnamefont {S.}~\bibnamefont
  {Sasanuma}},\ }\href@noop {} {\bibfield  {journal} {\bibinfo  {journal}
  {Brain and Language}\ }\textbf {\bibinfo {volume} {2}},\ \bibinfo {pages}
  {369} (\bibinfo {year} {1975})}\BibitemShut {NoStop}%
\bibitem [{\citenamefont {Eco}(1976)}]{ecooo}%
  \BibitemOpen
  \bibfield  {author} {\bibinfo {author} {\bibfnamefont {U.}~\bibnamefont
  {Eco}},\ }\href@noop {} {\emph {\bibinfo {title} {A Theory of Semiotics}}}\
  (\bibinfo  {publisher} {Indiana University Press},\ \bibinfo {address}
  {Bloomington, IN},\ \bibinfo {year} {1976})\BibitemShut {NoStop}%
\bibitem [{\citenamefont {Ryan}\ and\ \citenamefont {Schwartz}(1956)}]{ryans}%
  \BibitemOpen
  \bibfield  {author} {\bibinfo {author} {\bibfnamefont {T.~A.}\ \bibnamefont
  {Ryan}}\ and\ \bibinfo {author} {\bibfnamefont {C.~B.}\ \bibnamefont
  {Schwartz}},\ }\href@noop {} {\bibfield  {journal} {\bibinfo  {journal}
  {American Journal of Psychology}\ }\textbf {\bibinfo {volume} {96}},\
  \bibinfo {pages} {66} (\bibinfo {year} {1956})}\BibitemShut {NoStop}%
\bibitem [{\citenamefont {Gruber}\ and\ \citenamefont
  {B\"odeker}(2005)}]{gruberbook}%
  \BibitemOpen
  \bibinfo {editor} {\bibfnamefont {H.~E.}\ \bibnamefont {Gruber}}\ and\
  \bibinfo {editor} {\bibfnamefont {K.}~\bibnamefont {B\"odeker}},\ eds.,\
  \href@noop {} {\emph {\bibinfo {title} {Creativity, Psichology and the
  History of Science}}}\ (\bibinfo  {publisher} {Springer},\ \bibinfo {year}
  {2005})\BibitemShut {NoStop}%
\bibitem [{not({\natexlab{a}})}]{note1}%
  \BibitemOpen
  \bibinfo {note} {E.g. the \emph{General Conference on Weights and Measures},
  see {\tt www.bipm.org}, {\tt
  www.bipm.org/utils/common/pdf/CGPM/CGPM25.pdf}.}\BibitemShut {Stop}%
\bibitem [{not({\natexlab{b}})}]{note2}%
  \BibitemOpen
  \bibinfo {note} {Variable $\gamma$ is frequently a summation index which can
  represent a wide variety of variables; $H^0$ may be misunderstood as a
  Hamiltonian; $G$ can represent the gravity constant, among many other
  quantities whose names begin with a 'G'.}\BibitemShut {Stop}%
\end{thebibliography}%


\end{document}